\documentclass[iop]{emulateapj}

\newcommand{\HST}{{\it HST}}
\newcommand{\kms}{\ifmmode {\rm km\ s}^{-1} \else km s$^{-1}$\fi}
\newcommand{\Msun}{\ifmmode {\rm M}_{\odot} \else M$_{\odot}$\fi}
\newcommand{\Lsun}{\ifmmode {\rm L}_{\odot} \else L$_{\odot}$\fi}
\newcommand{\qo}{\ifmmode q_{\rm o} \else $q_{\rm o}$\fi}
\newcommand{\Ho}{\ifmmode H_{\rm o} \else $H_{\rm o}$\fi}
\newcommand{\ho}{\ifmmode h_{\rm o} \else $h_{\rm o}$\fi}

\newcommand{\vFWHM}{\ifmmode v_{\mbox{\tiny FWHM}} \else
                    $v_{\mbox{\tiny FWHM}}$\fi}
\newcommand{\CCF}{\ifmmode F_{\it CCF} \else $F_{\it CCF}$\fi}
\newcommand{\ACF}{\ifmmode F_{\it ACF} \else $F_{\it ACF}$\fi}
\newcommand{\Halpha}{\ifmmode {\rm H}\alpha \else H$\alpha$\fi}
\newcommand{\Hbeta}{\ifmmode {\rm H}\beta \else H$\beta$\fi}
\newcommand{\Hgamma}{\ifmmode {\rm H}\gamma \else H$\gamma$\fi}
\newcommand{\Hdelta}{\ifmmode {\rm H}\delta \else H$\delta$\fi}
\newcommand{\Lya}{\ifmmode {\rm Ly}\alpha \else Ly$\alpha$\fi}
\newcommand{\Lyb}{\ifmmode {\rm Ly}\beta \else Ly$\beta$\fi}
\newcommand{\HeI}{\ifmmode {\rm He}\,{\sc i}\,\lambda5876 \else 
	          He\,{\sc i}\,$\lambda5876$\fi}
\newcommand{\HeII}{\ifmmode {\rm He}\,{\sc ii}\,\lambda4686 \else 
	           He\,{\sc ii}\,$\lambda4686$\fi}

\newcommand{\heii}{He\,{\sc ii}}

\newcommand{\ciii}{\ifmmode {\rm C}\,{\sc iii} \else C\,{\sc iii}\fi}
\newcommand{\civ}{\ifmmode {\rm C}\,{\sc iv} \else C\,{\sc iv}\fi}
\newcommand{\CIV}{\ifmmode {\rm C}\,{\sc iv}\,\lambda1549 \else 
	           C\,{\sc iv}\,$\lambda1549$\fi}

\newcommand{\nv}{N\,{\sc v}}
\newcommand{\oi}{O\,{\sc i}}

\newcommand{\oiii}{O\,{\sc iii}}
\newcommand{\ob}{[O\,{\sc iii}]\,$\lambda \lambda 4959,5007$}

\newcommand{\mgii}{Mg\,{\sc ii}}

\shorttitle{Mrk\,590 Changes Type}
\shortauthors{}

\accepted{2014 September 10}

\begin{document}

\title{AGN Type-casting:  Mrk\,590 No Longer Fits the Role}

\author{ K.~D.~Denney\altaffilmark{1,5}, G.~De~Rosa\altaffilmark{1,2},
  K.~Croxall\altaffilmark{1}, A.~Gupta\altaffilmark{1,3}, M.~C.~Bentz\altaffilmark{4}, M.~M.~Fausnaugh\altaffilmark{1}, C.~J.~Grier\altaffilmark{1}, P.~Martini\altaffilmark{1,2}, S.~Mathur\altaffilmark{1,2}, B.~M.~Peterson\altaffilmark{1,2}, R.~W.~Pogge\altaffilmark{1,2}, B.~J.~Shappee\altaffilmark{1}}

\altaffiltext{1}{Department of Astronomy, 
		The Ohio State University, 
		140 West 18th Avenue, 
		Columbus, OH 43210, USA;
		denney@astronomy.ohio-state.edu}
		
\altaffiltext{2}{Center for Cosmology and AstroParticle Physics, 
                 The Ohio State University,
		 191 West Woodruff Avenue, 
		 Columbus, OH 43210, USA}

\altaffiltext{3}{Department of Biological and Physical Sciences, 
                      Columbus State Community College, 
                      Columbus, OH 43215, USA}

\altaffiltext{4}{Department of Physics and Astronomy,
		 Georgia State University,
		 Atlanta, GA 30303, USA}


\footnotetext[5]{NSF Astronomy \& Astrophysics Postdoctoral Fellow}

\begin{abstract}
  We present multi-wavelength observations that trace more than 40 years in the life of the active galactic nucleus (AGN) in Mrk\,590, traditionally known as a classic Seyfert 1 galaxy.  From spectra recently obtained from \HST, {\it Chandra}, and the Large Binocular Telescope, we find that the activity in the nucleus of Mrk\,590 has diminished so significantly that the continuum luminosity is a factor of 100 lower than the peak luminosity probed by our long baseline observations.  Furthermore, the broad emission lines, once prominent in the UV/optical spectrum, have all but disappeared. Since AGN type is defined by the presence of broad emission lines in the optical spectrum, our observations demonstrate that Mrk\,590 has now become a ``changing look'' AGN.  If classified by recent optical spectra, Mrk\,590 would be a Seyfert $\sim$1.9$-$2, where the only broad emission line still visible in the optical spectrum is a weak component of \Halpha.  As an additional consequence of this change, we have definitively detected UV narrow-line components in a Type 1 AGN, allowing an analysis of these emission-line components with high-resolution COS spectra.  These observations challenge the historical paradigm that AGN type is only a consequence of the line of sight viewing angle toward the nucleus in the presence of a geometrically-flattened, obscuring medium (i.e., the torus).  Our data instead suggest that the current state of Mrk\, 590 is a consequence of the change in luminosity, which implies the black hole accretion rate has significantly decreased.     
\end{abstract}




\section{INTRODUCTION}

The dichotomy separating active galactic nuclei (AGN) into Type 1 --- those observed to have broad emission lines --- and Type 2 --- those without broad lines --- originated from the first observations of \citet{Seyfert43} that demonstrated that the nebular lines in the nuclei of these nearby ``Seyfert'' galaxies sometimes had broad emission wings superposed with a narrow core and sometimes did not.  A working definition of Type 1 and Type 2, based on the relative line widths of the forbidden and Balmer lines was then elucidated by the work of \citet{Khachikian&Weedman74}.  \citet{Osterbrock&Koski76} and \citet{Osterbrock77, Osterbrock81} later note that the observed emission spectra of Seyfert galaxies are not so simple and argue for a continuum of intermediate types between these two extremes, e.g., Seyfert 1.2, 1.5, 1.8, and 1.9 based on the relative strength of the narrow \Hbeta\ component with respect to the broad component.  \citet{Antonucci85} presented a plausible scenario to physically explain this dichotomy by invoking different line of sight orientations:  the broad line-emitting region (BLR) in Type 2 AGN (Sy2s) is obscured by an optically thick, geometrically flattened medium (often referred to as the torus) leading to a ``hidden" BLR (HBLR), whereas the BLR is visible in Type 1 AGN (Sy1s) because our line of sight is not obscured by this optically thick medium.  This unified model of AGN \citep[see also][]{Antonucci93} followed from the discovery by \citet{Antonucci85} that broad emission lines were present in the polarization spectrum of NGC\,1068. The key assumption in this model is that all AGN systems are physically similar, with generally disk-like geometries. Thus, it is only the line of sight orientation of the BLR with respect to the obscuring medium that results in the observed type differences.

The unified model for Sy1 and Sy2 systems has been challenged by observations that suggest that not {\it all} AGN conform to this paradigm. Sy2s that do not have broad lines observed even through spectropolarimetry have been coined non-HBLR or ``true" Sy2s, and some may, in fact, be ``bare" Sy2s that are postulated to be devoid of the typical obscuring medium surrounding the BLR \citep[see e.g.,][]{Barth99, Tran01, Tran03, Panessa&Bassani02, Laor03, Zhang&Wang06}.  On the other hand, \citet{Antonucci12} advises caution before classifying objects as non-HBLR objects, as finding reflected broad lines is largely serendipitous and dependent on the particular source having a well-placed scattering medium with the right properties. Thus, the existence of true or bare Sy2's may not be well quantified.  Nonetheless, there are also objects known as ``changing look" AGN, which are more easily identifiable because an obvious change has occurred in the observed spectrum to warrant the application of this classification.  This characterization was originally coined based on X-ray observations in which sources appear alternately Compton-thin or ``reflection-dominated" (likely Compton thick) over the course of years \citep[e.g.,][]{Bianchi05}.   This qualifier has recently been extended to include objects which sometimes appear to have Sy2 and sometimes Sy1 characteristics in their optical spectrum \citep[e.g.,][]{Shappee13}.  Several examples of these optical changing look AGNs appear the literature over the years.  One of the most well-cited early examples is that of NGC\,4151, one of Seyfert's original galaxies assigned Type 1.5 by \citet{Osterbrock77}, but in which the optical broad lines all but disappeared (except for weak and possibly asymmetric wings) in the 1980's \citep{Antonucci83, Lyutyj84, Penston84} and have since returned \citep[see, e.g.,][]{Shapovalova10}.  

There are two typically accepted postulates to explain AGN changing their type: (1) variable obscuration or (2) variable accretion rate.  Variations in the obscuring medium is more suited to the unification paradigm of type resulting from different viewing angles and is typically invoked to explain changing look AGN in the X-ray regime \citep[e.g.,][]{Bianchi05, Risaliti09, Marchese12, Marin13}.  In contrast, it has been suggested that variations in luminosity and accretion rate are not only responsible for type changes in individual objects but are also responsible for the whole AGN typing sequence. In other words, the structure of the BLR changes with accretion rate, and objects evolve from high accretion rate when the AGN turns on --- Type 1 --- to low accretion rate once the AGN has depleted its nuclear material --- Type 2 \citep{Tran03, Wang&Zhang07, Elitzur14}, possibly oscillating between Type 1 and Type 2 and/or intermediate types between high and low accretion states while the nucleus remains active \citep{Penston84, Korista&Goad04}.  At sufficiently low accretion rates, many postulates exist demonstrating that a radiatively efficient BLR simply cannot be supported, due to, e.g., (1) a dearth of ionizing photon flux \citep{Korista&Goad04}, (2) the critical radius at which the accretion disk changes from gas pressure-dominated to radiation pressure-dominated becoming smaller than the inner-most stable orbit \citep{Nicastro00, Nicastro03}, (3) mass conservation considerations in the paradigm that the BLR arises in a disk wind with a fixed radial column, where the mass outflow rate cannot be sustained \citep{Elitzur&Ho09}, or (4) the accretion disk structure changing, replacing the disk-wind BLR with a radiatively inefficient accretion flow (RIAF) consisting of a fully ionized, low-density plasma incapable of producing broad lines \citep{Trump11}.  \citet{Laor03} similarly suggests a bolometric luminosity below which the BLR cannot exist, though with a slightly different argument following from a maximally observed broad line width.  

Observational results suggest that both of these physical processes are likely at play in different objects.  For example, \citet{Alexander13} present a serendipitously detected {\it NuSTAR} source at redshift $z=0.510$ for which observations suggest that significant obscuration may have moved into our line of sight to the nucleus of this source, changing it from a Type 1 to Type 2:  SED template fitting \citep[following][]{Assef10} show its SED is currently consistent with other optically identified Type 2 AGN in the same sample and an $E(B-V) = 0.6 \pm$0.5 mag, yet based on archival observations, its SED was previously consistent with that of a Type 1 AGN with estimated $E(B-V) = 0.00\pm$0.01 mag.  \citet{Alexander13} also presents a recent optical spectrum of this source, J183443+3237.8, that shows narrow emission lines and significant reddening of the continuum, though there is evidence for a weak broad component of \mgii.  On the other hand, \citet{Shappee13atel} report on the interesting behavior of NGC\,2617, which has changed from a Sy1.8 \citep{Moran96} to a Sy1, likely (though not conclusively) due to a recent outburst that triggered a transient source alert within the All-Sky Automated Survey for SuperNovae (ASAS-SN\footnote{http://www.astronomy.ohio-state.edu/$\sim$assassin}).  Subsequent observations of this AGN demonstrated a continued increase of the optical through X-ray emission by about an order of magnitude compared to past observations \citet{Shappee13}.  Due to the outburst nature of this activity, which occurred over relatively short time scales, it is highly unlikely that obscuration moving out of the line of sight could be responsible for this change in type.   Additionally, the optical spectral changes between Type 2 and Type 1.9 of NGC\,2992 also seem to be at least loosely correlated with large variability amplitudes (factors of a few tens) observed in the X-rays over both short (year) and long (decades) time scales, such that broad \Halpha\ only seems to be observed coincident with high X-ray states \citep[see][]{Gilli00, Murphy07, Trippe08}.

In this work, we examine the interesting and extreme phenomenon of Mrk\,590 (alt.\ NGC\,863) --- a classic Sy1 galaxy \citep{Osterbrock77, Weedman77} at $z=0.026385$ --- which appears to have changed from Type 1.5 to Type 1 then to Type $\sim$1.9$-$2, based on chronologically catalogued multi-wavelength observations.  Observations from MDM Observatory in late 2012 revealed that the broad Balmer emission lines seem to have disappeared from Mrk\,590.  Observations with higher S/N obtained in 2013 February with the MODS1 spectrograph on the Large Binocular Telescope confirmed that there was no longer any broad component to \Hbeta\ and possibly only a very weak broad \Halpha\ component.  Because this system is so well-studied, we have  been able to gain additional understanding of this change through multi-wavelength observations taken in its past and present states.  In Section \ref{S_data} we describe the new and archival data we have gathered.  We present our optical continuum fitting methods in Section \ref{S_hostfit} and discuss the overall observed trends in Section \ref{S_40yearprop}.  Section \ref{S_discuss} follows with a discussion of the properties of the nuclear region of Mrk\,590 as well as potential implications of the observed behavior of Mrk\,590.  We summarize our findings and plans for future work in Section \ref{S_summary}.  A cosmology with $\Omega_{m}=0.3$, $\Omega_{\Lambda}=0.70$, and $H_0 = 70$ km sec$^{-1}$ Mpc$^{-1}$ is assumed where necessary.

\section{New and Archival Observations}
\label{S_data}

We have collected both archival and new data covering the X-ray, UV, and optical wavelength regimes in an attempt to better understand the recent behavior of Mrk\,590.

\subsection{Optical Data}

We present a selection of 10 optical spectra that span more than four decades of observations of Mrk\, 590.  These have been gleaned from the personal archives of the authors, the literature, the Sloan Digital Sky Survey (SDSS), and new observations.  Data are presented in chronological order of the known or presumed date of observation:
\begin{enumerate}
\item Historical Lick Observatory spectra of Mrk\,590 from 1973 were measured from copies of the original reduced spectra provided by the late Prof.\ Donald Osterbrock to one of the authors (RWP). Details of the observation and reduction of these spectra are given in \citet{Osterbrock77}, however, we note two important items about these spectra.  First, no explicit date is available, either in the spectral image headers (which were converted from raw IDS output to FITS format in 1996), or in the \citet{Osterbrock77} paper.  Second, \citet{Osterbrock77} states that all of the spectra were taken ``in the years 1974$-$1976", so it is possible that these spectra are from 1974, though it is equally likely that 1973 is correct, and they were taken early during the commissioning of the IDS in 1972/73.  The entrance aperture of this spectrograph was 2\farcs7 $\times$ 4\arcsec.

\item Spectra of Mrk\,590 were obtained at the 1.8-m Perkins telescope at Lowell Observatory with the Ohio State University Image Dissector Scanner \citep{Byard81} through a 7\arcsec\ round aperture in 1984$-$1986 and have been previously discussed in the literature by \citet{Ferland90}.  Here, we present spectra from JD2445672 and JD2445674 --- 1983 December --- which cover the blue (\Hbeta) and red (\Halpha) sides of the spectrum, respectively. We chose these spectra from the full set because of the near simultaneity of the red and blue wavelength coverage, fewer relative flux calibration issues, and/or the longer combined wavelength coverage than afforded by other spectra in this set.  We merged these nearly contemporaneous spectra, multiplicatively scaling the red spectrum by a factor of 1.08 to match the blue spectrum. 

\item Mrk\,590 was targeted in an early reverberation mapping campaign, during which 102 spectra were obtained of the \Hbeta\ emission region between JD2447837 and JD2450388 (1989 Nov -- 1996 Oct) and from which the black hole mass was determined to be $(4.75 \pm 0.74) \times 10^7M_\odot$ \citep{Peterson04}.  Details of the campaign and observations are described by  \citet{Peterson98}. For completeness we note that the spectra were taken through a 5\arcsec\ slit and extracted through a 7\farcs6 aperture.  We took spectra from three epochs during this campaign: JD2447837 --- 1989 November --- the beginning of the campaign, JD2448999 --- 1993 January --- chosen to be roughly 10 years after the previous 1983 spectrum, and JD2450365 --- 1996 Oct --- which was near the end of the campaign.  

\item Mrk\,590 was observed by the Sloan Digital Sky Survey (SDSS) on MJD2452649 --- 2003 January --- serendipitously continuing our roughly 10 year interval between archival optical spectra.  The SDSS spectra were obtained through a 3\arcsec\ diameter fiber and spectrophotometric calibration is based on simultaneously observed field stars. 

\item A spectrum of Mrk\,590 was obtained on the 1.3m McGraw-Hill telescope at MDM observatory on JD2454000 --- 2006 Sep --- using the CCD Spectrograph with the 350~l/mm grating.  The observation was obtained with a 1\arcsec slit and extracted through a 12\arcsec\ aperture. Unfortunately, all of the data obtained during this observing run suffer from the presence of an unexplained low-level and time-variable signal in the underlying bias levels that somewhat distorted the continuum shape of the Mrk\,590 spectrum.  Due to the time-variable nature of the signal, we were unable to correct the spectra, but we truncated the most blueward and redward wavelength ranges that were the most affected.  As a result, the flux calibration of the overall spectrum, and even that as a function of wavelength, is not highly reliable.  Nonetheless, we did not deem this issue problematic enough to justify omitting this spectrum, as it represents an important epoch in the changing nature of this object, as discussed below.

\item We used the MODS1 spectrograph on the Large Binocular Telescope to obtain an optical spectrum of Mrk 590 on 2013 February 14.  We used both the red and blue channel of MODS1 with the dichroic to obtain wavelength coverage from 3200$-$10000\,\AA. The blue-channel spectra used the G400L grating (400\,lines\,mm$^{-1}$ in first order), and the red-channel spectra used the G670L grating (250\,lines\,mm$^{-1}$ in first order). For all spectra we used the 1\farcs2 segmented long-slit mask (LS5x60x1.2), oriented along the parallactic angle to minimize the effects of differential atmospheric refraction.  The spectra were extracted with a 4\arcsec\ aperture.

\item We next observed Mrk\,590 on 2013 December 18 during commissioning of the new Kitt Peak Ohio State Multi-Object Spectrograph (KOSMOS) on the Mayall 4-m telescope.  Observations were taken through an 0\farcs9 longslit, using both the red and blue VPH grisms to cover the wavelength range $\sim$4000$-$9000\,\AA.  The spectra were extracted with an 11\farcs6 aperture. 

\item Most recently, on 2014 January 7, another spectrum was taken on the 1.3m McGraw-Hill telescope at MDM observatory with the CCD Spectrograph and the 350 l/mm grating.  This time, the spectrum was taken through a 5\arcsec\ slit and extracted with a 15\arcsec\ aperture because it was observed at the start of a reverberation mapping program that requires these large apertures.

\end{enumerate}

\begin{figure*}
\epsscale{1.2}
\plotone{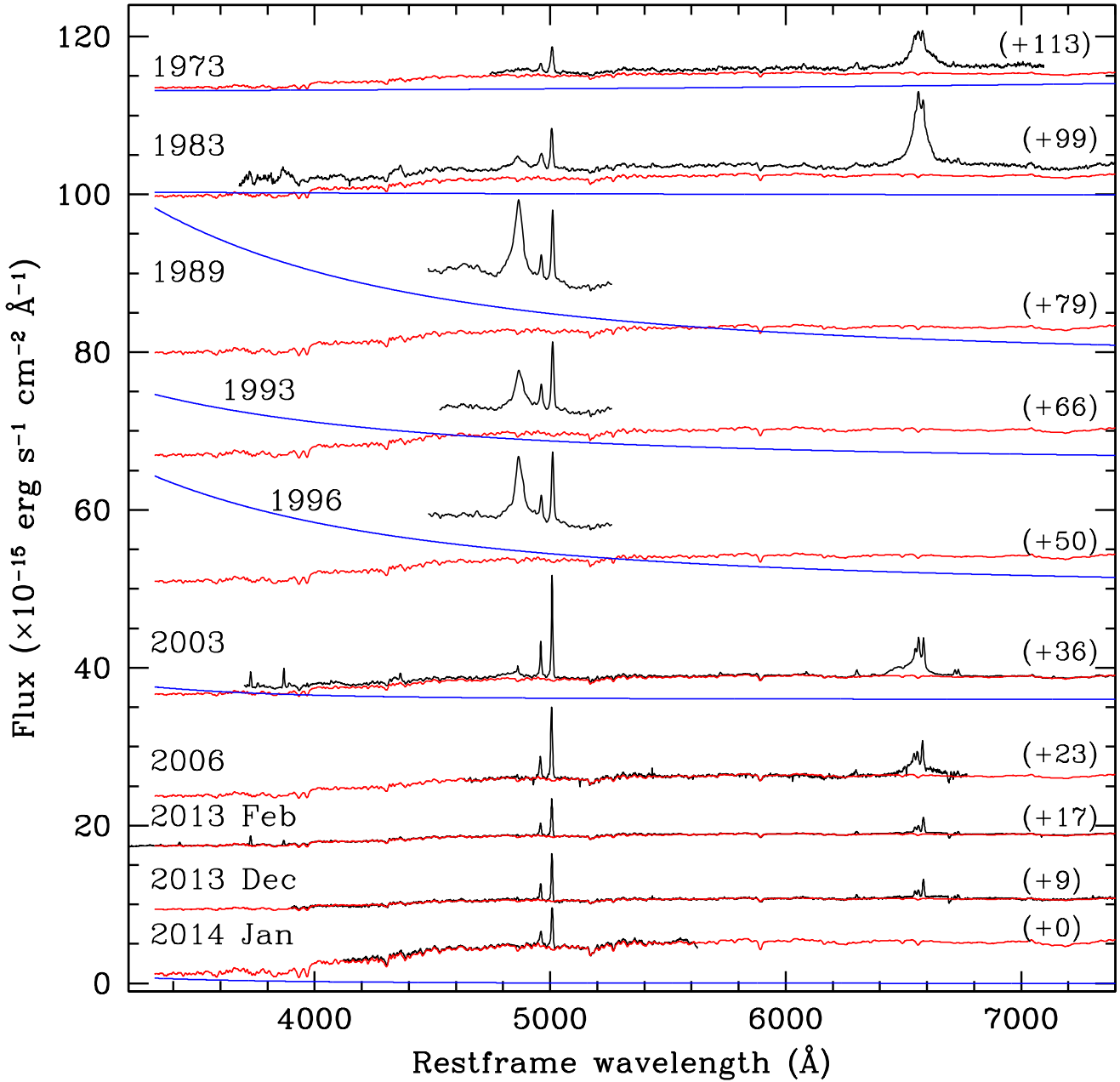}

\caption{Rest frame optical spectra of Mrk\,590 spanning more than 40 years.  The black curves show the observed spectra, the red curves show a host galaxy starlight template that was fit to the 2013 MODS1 spectrum and has been scaled to all other spectra (see Section \ref{S_hostfit}), and the blue curves show power law continuum fits for the epochs in which the stellar component could not account for all the observed continuum flux.  All spectra are on the same flux scale, given in the ordinate axis label, but a constant has been added (given in parentheses to the right of each spectrum) to all but the most recent epoch to clearly visualize observed changes between epochs.} 

\label{fig_allspec}
\end{figure*}

Figure \ref{fig_allspec} shows all 10 of the optical spectra discussed above.  These spectra were de-redshifted to the restframe of Mrk\,590 and resampled onto separate linear wavelength scales similar to the natural dispersion of each respective spectrum near 5100\,\AA.  The absolute spectrophotometry of several of these observations is not strictly reliable due to the uncertain conditions or known poor conditions under which they were taken.  The spectra with the most reliable absolute flux calibration are (1) the 1989, 1993, and 1996 RM campaign spectra, which were calibrated to the assumed constant [\oiii]\,$\lambda$5007 line flux measured from data taken under photometric conditions \citep[see][for details]{Peterson98}, (2) the 2003 SDSS spectrum, for which the absolute spectrophotometry is good to $\sim4$\%, and (3) the most recent MDM spectrum from 2014 January, which was taken along with a spectrophotometric standard star under clear conditions. 

\subsection{UV Data}

\begin{figure*}
\epsscale{1.2}
\plotone{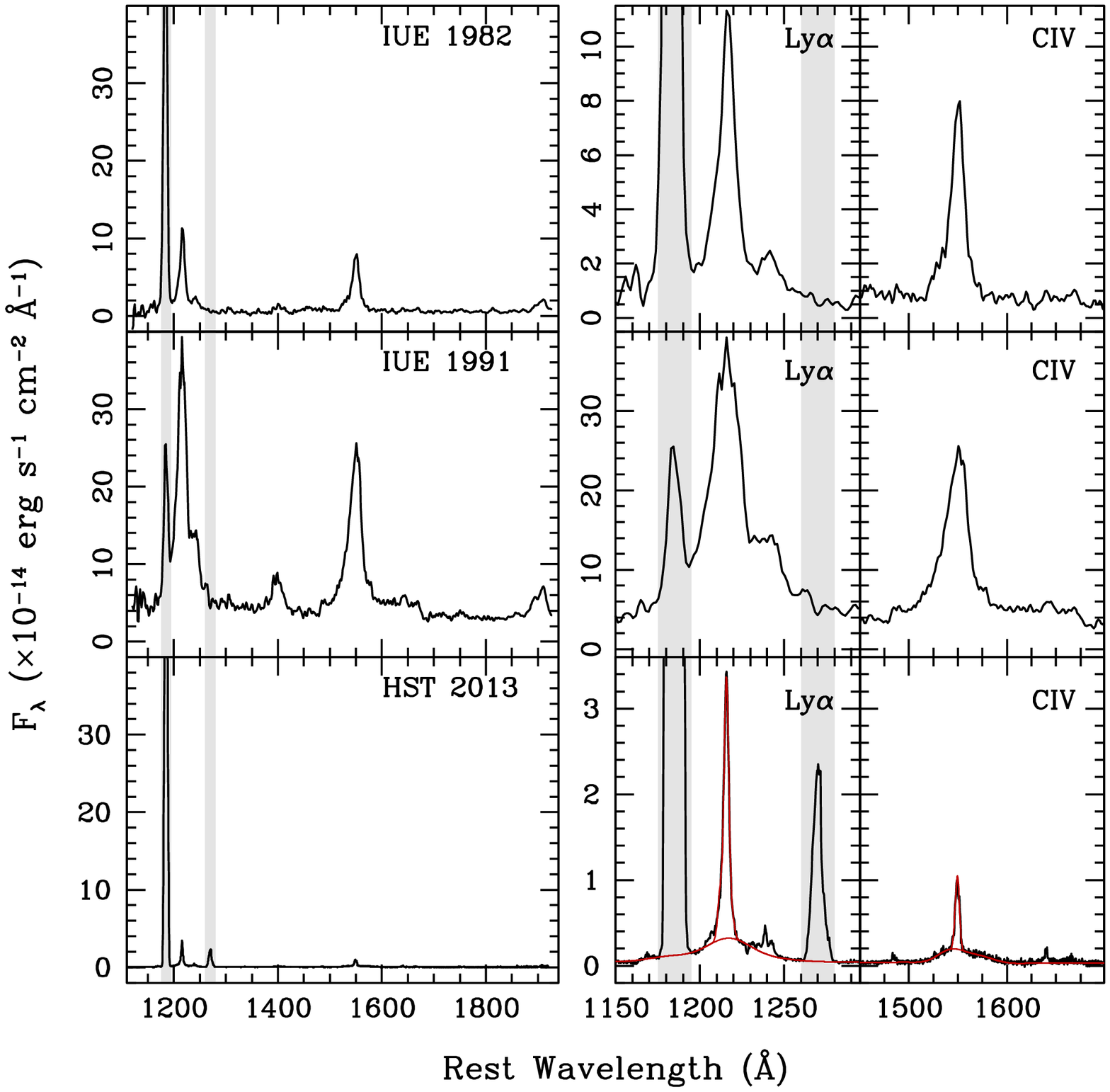}

\caption{Rest frame UV spectra of Mrk\,590.  The left panels show the full spectra from the available {\it IUE} spectra and new COS observations, with the year of the observation included with the source in the upper corner of each panel.  The flux units and ordinate range is the same for all spectra to demonstrate the factor of $\sim$100 change in luminosity of Mrk\,590 across these epochs. The middle and right panels in each row are zoomed in on the \Lya\ and \civ\ emission lines, respectively.  The red curves in the {\it HST} \Lya\ and \civ\ panels are functional fits to both the broad component and total line flux (see Section \ref{S_emissionlines} for details), and the gray shaded regions show the location of geocoronal \Lya\ and \oi\ emission from the Earth's atmosphere.}

\label{fig_uvspec}
\end{figure*}

After discovering the disappearance of the optical broad emission lines from the 2013 MODS1 spectrum, we obtained Director's Discretionary time (PID 13185) on the {\it Hubble Space Telescope (HST)} to investigate how this change affected the UV continuum and emission lines.  We obtained this new \HST\ observation of Mrk\,590 on 2013 July 17 with the G140L mode of the FUV detector on the Cosmic Origins Spectrograph (COS).  The data were processed with the standard pipelines, and the final 1D spectrum was de-redshifted and binned to a linear dispersion of 0.4\,\AA\,pixel$^{-1}$.  Mrk\,590 was previously observed in the UV by the {\it International Ultraviolet Explorer (IUE)} on 1982 July 20 and 1991 January 14.  We downloaded these spectra from the {\it IUE} archives and converted them to the multispec FITS format in IRAF.  We only analyzed the SWP channel spectra, which cover a similar wavelength range as the new COS observations, although LWP observations from both dates also exist.  We de-redshifted the IUE spectra to rest frame wavelengths and resampled them onto a linear wavelength scale of 1.5\,\AA\,pixel$^{-1}$.  These spectra are shown in the left panels of Figure \ref{fig_uvspec}.

\subsection{X-ray Data}

\begin{figure}
\epsscale{1.2}
\plotone{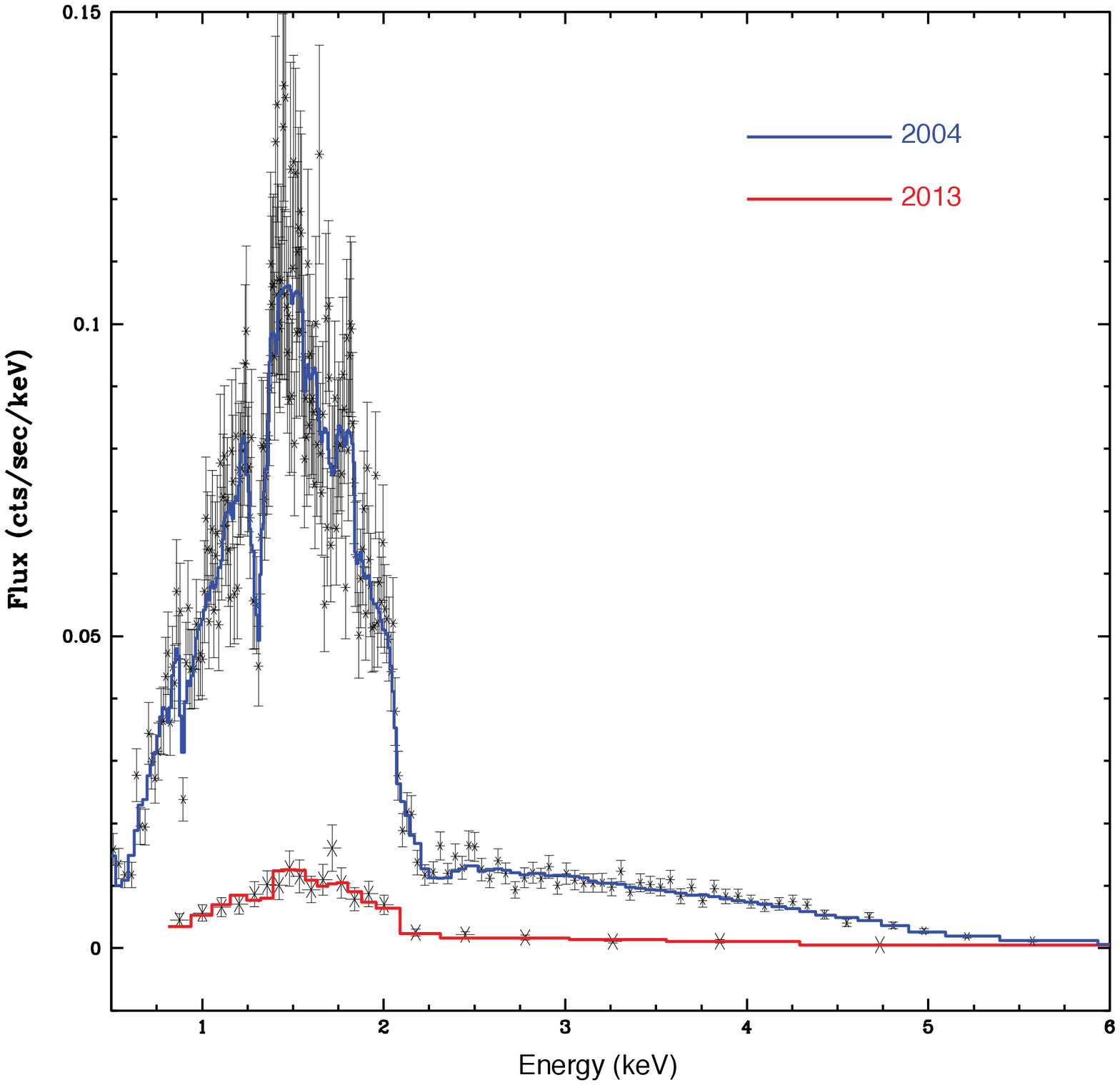}

\caption{Chandra observations of Mrk\,590.  The top spectrum is based on archival data from 2004, while the bottom spectrum is from the new observations described in this work.  The black points with error bars show the observations and the colored curves are absorbed power-law fits to the observed spectrum.} 

\label{fig_xrayspec}
\end{figure}

We also obtained {\it Chandra} Director's Discretionary time
contemporaneous with the new \HST\ observations, and thereby observed
Mrk\,590 with the HETGS on 2013 June 17 for $\sim$30\,ks to investigate
the consequences of the observed change in Mrk\,590 on the X-ray
spectrum.  The HETGS consists of two grating assemblies, a high-energy
grating (HEG) and a medium-energy grating (MEG). The HEG (MEG) bandpass
covers 0.8$-$10\,keV (0.5$-$10\,keV), but the effective area of both
instruments falls off rapidly at either end of the bandpass. We reduced
the data for both gratings using the standard Chandra Interactive
Analysis of Observations (CIAO) software (v4.3) and Chandra Calibration
Database (CALDB, v4.4.2) and followed the standard {\it Chandra} data
reduction
threads\footnote{\url{http://cxc.harvard.edu/ciao/threads/index.html}}. To
increase S/N we co-added the negative and positive first-order spectra
and built the effective area files (ARFs) using the \emph{fullgarf} CIAO
script.  We then binned the MEG spectrum (at least 20 counts per
bin).  Additionally, we analysed the zeroth order spectrum which has
somewhat higher S/N; there are 620 total counts (0.2--8 keV) in the
zeroth order, compared to 450 counts in the combined 1st order of MEG.
Nonethess, we plot the MEG spectrum in Figure \ref{fig_xrayspec} to show
how these new data compare with previous {\it Chandra} archival HETGS
observations from 2004 \citep{Longinotti07} that were reprocessed here
in a similar manner.

\section{Host Galaxy Template and Optical AGN Power-law Continuum Fitting}
\label{S_hostfit}

The active nucleus of Mrk\,590 sits within a nearby, typical SA(s)a galaxy.  As a result a significant portion of the light entering our spectroscopic apertures is host galaxy starlight rather than emission from the AGN itself.  This contamination must be subtracted to accurately assess the changing luminosity of the AGN.  \citet{Bentz09rl, Bentz13} used {\it HST} images to measure the host galaxy starlight contribution to the observed flux at rest frame 5100\,\AA, but the amount of host flux will depend on the spectroscopic aperture.  Bentz et al.\ report the flux within the spectroscopic aperture used for the \citet{Peterson98} RM campaign spectra of Mrk\,590 to be $(3.965\pm0.198) \times 10^{-15}$ erg s$^{-1}$ cm$^{-2}$ \AA$^{-1}$.  

The optical spectra discussed here were obtained through a variety of entrance apertures, so we estimate the starlight flux by fitting a starlight model to the 2013 MODS1 spectrum, chosen for its combination of long wavelength coverage, high spectral resolution ($\lesssim3$\,\AA), and paucity of thermal AGN continuum emission.   We modeled the underlying stellar continuum using the STARLIGHT\footnote{www.starlight.ufsc.br} spectral synthesis code \citep{starlight}. We have fit \citet{Bruzual03} models to the continuum between 3320 -- 9500\,\AA, using a \citet{Calzetti00} reddening law and masking nebular emission lines and the location of the Meinel bands where sky subtraction is challenging \citep[$\sim$7500\,\AA,][]{Meinel50}.  The intrinsic dispersion of the STARLIGHT model fit is 1\,\AA\,pixel$^{-1}$, so we resampled the model to the dispersion of each of our sample optical spectra and then matched the resampled model to each by applying a multiplicative scale factor to account for the different entrance apertures.  The resulting scale factors roughly follow expectations based on the aperture differences between spectra, though a strict scaling vs.\ aperture relation is not expected because of seeing effects and because the host galaxy starlight is not symmetric about the nucleus \citep[see][]{Bentz09rl}.  This method leads to an estimate of the starlight contribution near 5100\,\AA\ in the \citet{Peterson98} RM campaign spectra considered here from 1989, 1993, and 1996, which were all taken through the same aperture, of $(3.7\pm0.2) \times 10^{-15}$ erg s$^{-1}$ cm$^{-2}$ \AA$^{-1}$, which is consistent with the \citet{Bentz13} measurement.

There is no evidence for any underlying AGN thermal continuum emission in the 2013 MODS1 spectrum.  However, for epochs where an additional power-law component from the AGN continuum was required to account for all of the observed emission, we first scaled the starlight model to match the strength of host galaxy stellar absorption features (e.g., Mg Ib and/or NaD), and then included an additional power-law component of the form $F_{{\rm pl},\lambda} = F_{\rm pl,0}(\lambda/\lambda_0)^{\alpha}$, where the flux and wavelength normalization, $F_{\rm pl,0}$ and $\lambda_0$, respectively, and the power-law slope, $\alpha$, were left as free parameters.  These host starlight and power-law continuum components, where applicable, are shown by the red and blue curves, respectively, in Figure \ref{fig_allspec}. 

\section{40 Years of Continuum and Emission Line Properties}
\label{S_40yearprop}

\subsection{X-ray, UV, and Optical Continuum Properties}
\label{S_continuum}

Dramatic changes are clearly visible in Figure \ref{fig_allspec} in the optical nuclear emission from Mrk\,590 over the last four decades.  As mentioned above, the relative strength of the composite starlight component needed to fit each spectrum is a strong function of the spectroscopic aperture, but otherwise, as expected, the appearance and strength of the stellar features have not changed over the time spanned by our observations, and the model that was fit to the 2013 February spectrum is a good match to the features seen at all other epochs.  

The same cannot be said of the AGN continuum contribution, which can be estimated from the power law component.  However, because of (1) the varying (and often short) wavelength range of the available data, (2) the degeneracy between the power-law and starlight components, and (3) the sometimes questionable accuracy of the flux calibration and response curve correction within this heterogeneous data set, the details of the power-law fit parameters are not individually reliable (and are therefore not provided).  This is particularly true of the weak power-law component to the most recent, 2014 January spectrum, where the necessity of this component is not statistically significant, even at the 1$\sigma$ level, given the S/N and the short wavelength coverage of the spectrum.  Nonetheless, the relative increase (and then decrease) in strength and steepness of the power-law slope as the broad lines strengthen (and then weaken) from the 1970s to 1990s (the 1990s to the 2000s) is qualitatively significant.    

\begin{figure}
\epsscale{1.2}
\plotone{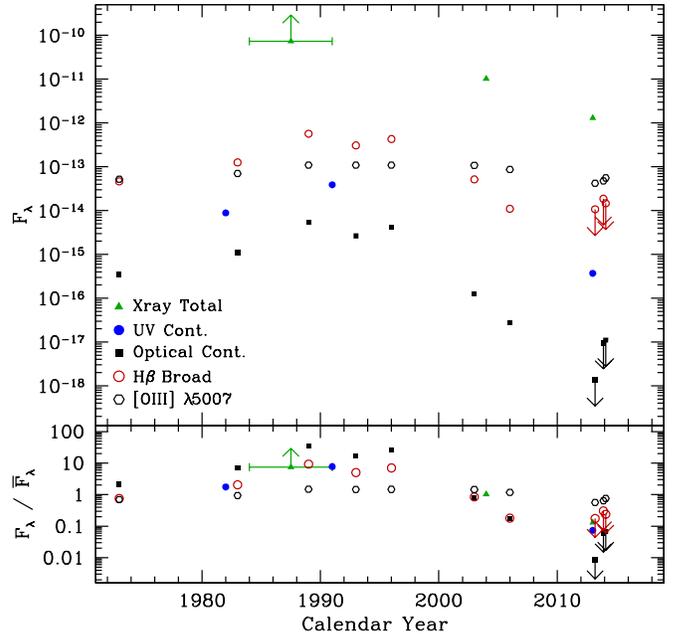}

\caption{Light curves showing the multi-wavelength changes observed in Mrk\,590 over the past four decades in the hard$+$soft X-ray flux (green triangles), UV continuum flux (filled blue circles), optical starlight-subtracted continuum flux (black squares), and \Hbeta\ and [\oiii]\,$\lambda$5007 emission line fluxes (open red and black points, respectively).  The top panel shows observed fluxes, where the UV and optical continuum fluxes are given in units of erg s$^{-1}$ cm$^{-2}$ \AA$^{-1}$, and the X-ray and emission-line fluxes are given in units of erg s$^{-1}$ cm$^{-2}$.  The bottom panel shows fluxes in the same units but relative to the geometric mean flux state of each emission component over the course of our observations, $\bar{F}_{\lambda}$.  Following from Table~1 and the discussion in Section \ref{S_40yearprop}, the earliest X-ray flux is taken to be a lower limit representing the sum of the 1984 EXOSAT and 1991 RASS measurements (the error bar represents the time spanning these observations), and the more recent optical continuum and \Hbeta\ emission line fluxes are upper limits.  Uncertainties on the other flux values are not given per Table Note 1b.} 

\label{fig_lightcurves}
\end{figure}

The rise and fall of the optical AGN continuum over the past four decades can be quantitatively evaluated from the 5100\,\AA\ AGN continuum flux estimates for each of the spectra in our sample, provided in Table~1 and shown in Figure \ref{fig_lightcurves}, based on the starlight-subtracted optical spectra.  For the 2013 and 2014 spectra, the estimated continuum fluxes are strict upper limits because the weak residuals left after subtraction of the starlight fit are as likely to be noise and the consequence of an imperfect stellar model as they are minimal contributions from the AGN. Nonetheless, this still gives an indication of the change in flux over this time.  Based on this sample, the peak in both the broad emission line strength and the continuum flux is observed from the 1989 spectrum.  However, work by \citet{Korista90} and \citet{Ferland90} indicate that the peak continuum state may have been as early as 1988, as they report the optical, non-stellar continuum flux at 5000\,\AA\ to be $6.81\times 10^{-15}$ erg s$^{-1}$ cm$^{-2}$ \AA$^{-1}$ from a spectrum taken on 1988 Sept 8, which is $\sim$20\% higher than that in our 1989 spectrum.  These estimates suggest a weakening in the strength of the AGN continuum component of 2$-$3 orders of magnitude since the late 1980's, but stronger constraints on the changes in the AGN SED can be set with shorter wavelength observations, where the host galaxy no longer contributes significantly to the measured flux.

The UV continuum flux measurements from the {\it IUE} and COS spectra, also presented in Table 1 and Figure \ref{fig_lightcurves}, are affected neither by aperture and seeing-dependent contributions from the host galaxy starlight nor by weather-related flux calibration issues (though instrument-related flux calibration uncertainties are still possible).  Sampling the ionizing flux responsible for producing the emission line flux is not possible due to Galactic absorption, but the NUV continuum is the closest proxy we have and can therefore provide insight into the changes in the ionizing continuum.  We find that the flux at 1450\,\AA\ followed the same trends as in the optical, demonstrating a significant, factor of 4$-$5, increase in flux between 1982 and 1991, followed by a decline in flux of nearly a factor of 100 between 1991 and 2013.  Interestingly, the slope of the UV continuum, at least over this relatively short wavelength range, has remained relatively constant, unlike the optical continuum slope, insofar as it can be judged from the power-law fits.

Similar trends are present in the new and archival X-ray data and are
visible in Figure~\ref{fig_lightcurves}.  We performed spectral fitting
of the {\it Chandra} data shown in Figure \ref{fig_xrayspec} using the
CIAO fitting package \emph{Sherpa}.  The intrinsic continuum of the
source is well-fit with an absorbed ($N_{H}=2.7\times10^{20}$~cm$^{-2}$;
Dickey \& Lockman 1990) power law. The 0.5$-$8 keV power-law slope
fit to our new observations is $\Gamma=$1.52$\pm$0.12 for the MEG
spectrum and $\Gamma=$1.53$\pm$0.14 for the zeroth order spectrum. These
are consistent with each other and with the value of
$\Gamma=$1.58$\pm$0.15 measured from the the high state spectrum from
2004 (both shown in Figure~\ref{fig_xrayspec}). The intrinsic absorption
in both cases is consistent with zero; we find intrinsic
$N_{H}=0.0^{+0.06}\times10^{20}$~cm$^{-2}$ for the MEG spectrum and
$N_{H}=0.0^{+0.08}\times10^{20}$~cm$^{-2}$ for the zeroth order
spectrum, with lower limit pegged at zero in both cases.  We measured
the flux to be $(1.3 \pm 0.3) \times 10^{-12}~$erg~s$^{-1}$~cm$^{-2}$
over 0.5$-$10\,keV.  This measured X-ray flux is lower by an order of
magnitude compared to 2004 {\it Chandra} \citep[][and also shown in
Figure \ref{fig_xrayspec}]{Longinotti07} and {\it XMM} observations
\citep{Longinotti07, Bianchi09}, the latter of which show a 2$-$10\,keV
flux similar to that measured with {\it Suzaku} observations as recently
as 2011 \citep{Rivers12}, though the soft-excess observed earlier
had disappeared during the {\it Suzaku} observations.  These 2004
observations, in turn, are nearly an order of magnitude lower than the
combined X-ray flux from (1) 0.1$-$2.4\,keV observations taken in 1991
as part of the ROSAT All-Sky Survey \citep[RASS;][]{Voges99, Mahony10}
and (2) 2$-$10\,keV EXOSAT observations from 1984 \citep{Turner89}.  We
take the combined RASS+EXOSAT flux shown in Figure~\ref{fig_lightcurves}
as a lower limit to the expected soft$+$hard flux for the epoch
$\sim$1990 (the point appears at the average calendar year and the
x-axis error bar spans the range between the two observations).  This is
because the optical and UV data suggest the continuum luminosity was
rising throughout the 1980s, so that the 1984 EXOSAT flux is likely
underestimated compared what may have been expected if observations had
been taken $\sim$6 years later.  All X-ray fluxes are summarized in
Table 1.  We have been also awarded an additional 70\,ks of {\it
Chandra} time to observe Mrk\,590 again later this Cycle, the results of
which will appear in a future contribution.

While we have not yet been able to locate spectra taken between the years 1996 and 2003, the MAGNUM telescope photometrically monitored the continuum variability of Mrk\,590 from 2001 Sept 11 -- 2007 Aug 8.  We can therefore use this and previous monitoring data \citep[particularly that of][]{Peterson98} to comment on the continuum variability trends over the last couple of decades.  The MAGNUM light curves are presented by \citet{Sakata10} and show that the amplitude of variability of Mrk\,590 was not as large as other typical Seyferts (e.g., NGC\,4151 and NGC\,5548) monitored by the same program.  Nonetheless, Mrk\,590 continued to exhibit variability typical of this AGN luminosity regime, i.e., relatively larger amplitude, secular variability over longer, inter-year time scales, as well as smaller amplitude, more rapid, intra-year variability patterns.   The maximum {\it B}-band flux over this period occurred around JD2453300 --- 2004 Oct --- and was similar to the flux at the start of the MAGNUM observations.  By the end of the MAGNUM observations, the broad \Hbeta\ emission had disappeared, as seen in Figure \ref{fig_allspec}, but the {\it B}-band flux was only $\sim$15\% lower than the maximum 2004 level, and the MAGNUM light curves show neither a steady decline nor even a leveling off of the continuum to a constant flux by the end of the monitoring period.  So despite the lack of broad lines and the relatively lower mean luminosity compared to when the broad lines were strong, the continuum variability behavior was relatively normal for a Sy1 of this luminosity and black hole mass.  Similarly, the continuum light curves from the reverberation mapping campaign of \citet{Peterson98} are not particularly enlightening with regard to the transformation of Mrk\,590, except that the AGN simply appears to be in a higher luminosity/accretion state over this time period. The observed variability is larger in amplitude than the MAGNUM campaign, but still well within typical for moderate luminosity AGN.  Furthermore, there are no apparent `flaring' events, such as that observed in NGC2617 \citep{Shappee13}, at least within the data we have collected.  It therefore is unlikely that this intriguing transformation of Mrk\,590 could easily have been gleaned simply from optical/NIR broadband monitoring, likely due to contamination from the host galaxy in the optical.

\subsection{Emission Line Properties}
\label{S_emissionlines}

\begin{figure*}
\epsscale{1.2}
\plotone{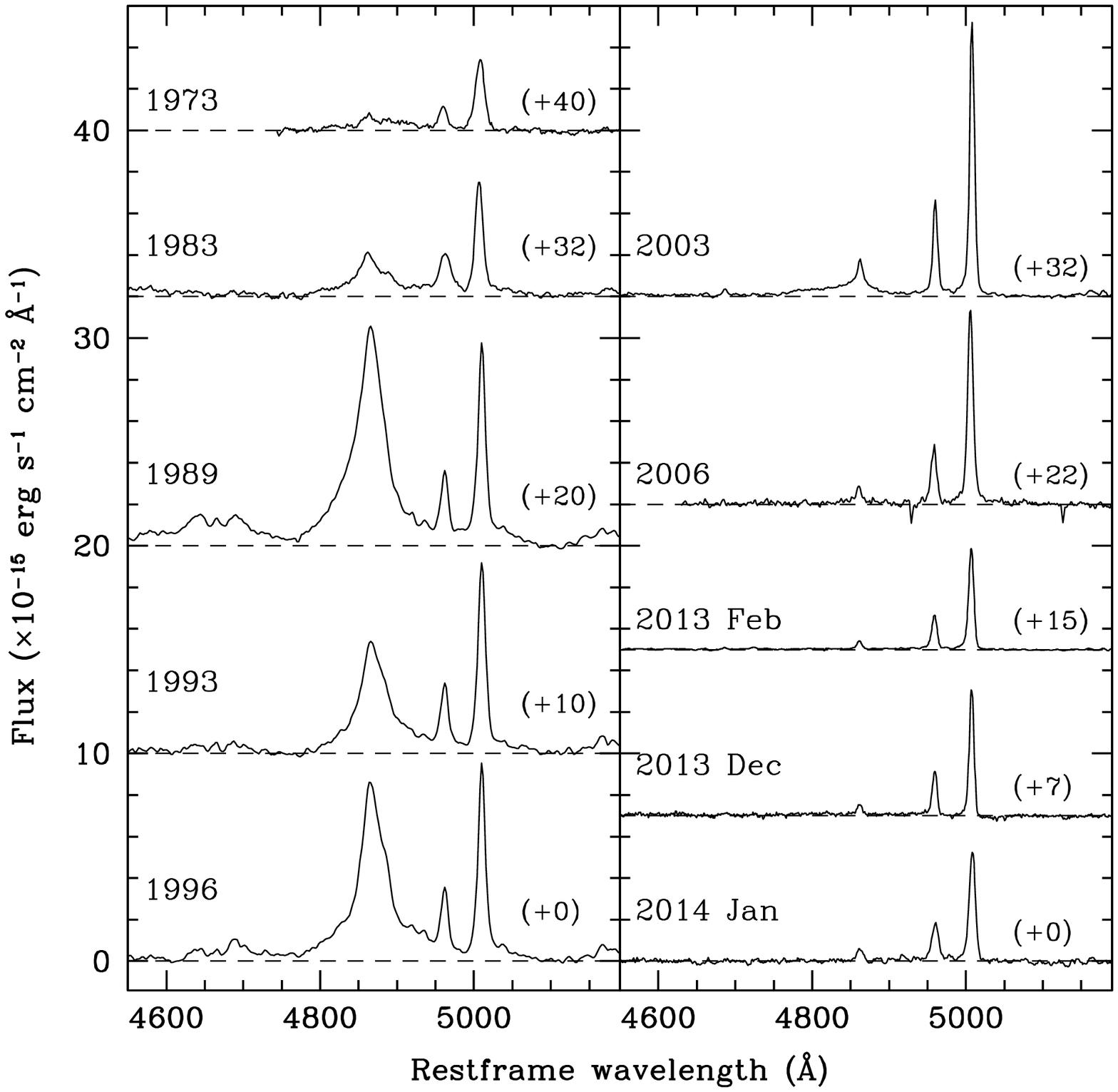}

\caption{The starlight and power-law continuum-subtracted, residual spectra of Mrk\,590, zoomed to exhibit only the \Hbeta\ and \ob\ emission-line region of the 10 optical spectra of Mrk\,590 from Figure \ref{fig_allspec}.  The year and/or month of observation is given to the left of each spectrum.  The dashed lines represent the artificial zero-level continuum under each spectrum after adding the flux offsets given in parentheses to the right of each spectrum.  The spectra are otherwise on the same flux scale.}

\label{fig_zoomspec}
\end{figure*}

Figure \ref{fig_zoomspec} shows a zoomed-in view of the \Hbeta\ and \ob\ emission lines from the starlight- and continuum- (where applicable) subtracted optical spectra presented in Figure \ref{fig_allspec}.  Significant changes in the emission-line properties are obvious in both the broad and narrow emission lines.  The most dramatic change is the initial strengthening and then complete disappearance of the broad \Hbeta\ line.  This trend is also visible in the narrow emission-line-subtracted \Hbeta\ light curve shown with the open red circles in Figure \ref{fig_lightcurves}, where the final 3 epochs shown as upper limits are from the recent MODS1, KOSMOS, and 2014 MDM spectra, where the broad \Hbeta\ line is no longer visible in the spectrum.  Generally speaking, broad emission-line profile changes are not particularly informative, given our current understanding of the BLR.  Significant changes are well-known to occur over long timescales and in objects that have not ``turned off'' as Mrk\,590 appears to have done \citep[see, e.g.,][]{Wanders&Peterson96, Shapovalova04, Shapovalova09}.  The flux of the broad lines also naturally varies over time, so, for example, the changes observed within the RM campaign spectra from 1989, 1993, and 1996 are typical for Sy1s.  However, the dramatic changes observed here, i.e., between 1973 and 1989 and again between 1996 and 2013, are likely connected to the dramatic changes occurring on similar timescales in the level of the AGN continuum flux.  These observed changes have been predicted from photoionization modeling of the BLR, which demonstrates that the equivalent widths and responsivity of the optical recombination lines vary with the incident continuum flux \citep{Korista&Goad04}.  Furthermore, \citet{Korista&Goad04} find that optical depth effects within the BLR predict larger responsivities for higher order Balmer lines compared to the lower order lines, with the effect being a steepening (flattening) of the broad-line Balmer decrement in low (high) continuum states.  These results predict the variation of the intermediate Seyfert type (1.5$-$1.9) over large changes in the continuum flux, as observed in these changing type Seyferts, such at NGC\,4151, NGC\,2617, possibly NGC\,2992, and here in Mrk\,590.

While the broad emission line flux changes are the most dramatic, the relative changes in the NLR emission lines are actually the most intriguing, as well as physically enlightening, and allow a possible history of the central region of Mrk\,590 to be at least partially assembled.  Narrow line fluxes are given in Table 1 and the [\oiii]\,$\lambda$5007 emission line light curve is shown in Figure \ref{fig_lightcurves}, both of which demonstrate that the [\oiii]\,$\lambda$5007 flux varied significantly across these four decades.  Reverberation mapping studies have shown that while the BLR emission is variable on relatively short timescales --- hours to weeks --- the NLR emission remains constant over much longer timescales because of the radial distance from the central source to the NLR, geometric damping or smoothing of any variability due to the large spatial extent of the NLR, and long recombination times due to the low densities in the NLR.  The NLR will not ``reverberate'' in response to small changes in the ionizing continuum emission as the BLR does.  Narrow-line variability has only been previously noted in the literature a couple of times.  \citet{Antonucci84b} report changes in narrow lines flux and flux ratios, but only for narrow line radio galaxies, and \citet{Ferland79}, \citet{Zheng95}, and \citet{Zheng96} measure changes in the narrow-line fluxes in radio-loud AGN 3C\,390.3.  In the case of the latter object, the narrow line fluxes are observed to decline within only $\sim\,4$ months of the continuum decline, and \citet{Zheng95} and \citet{Zheng96} suggest the narrow-line emission in this source may not only be exceptionally compact and dense, but also be affected by the jet, which is known to produce a beamed component of the continuum emission.  

It was demonstrated only recently by \citet{Peterson13} that the NLR emission in a radio quiet AGN, NGC\,5548, does vary over long time scales, i.e., a few to tens of years, and responds to large changes in the ionizing continuum with a damping, or smoothing, time scale of $\sim$\,15 years for this object.   We observe similar changes in the narrow-line flux in Mrk\,590.  Our absolute flux calibration is admittedly somewhat uncertain for many of our optical spectra, but the general trend seen and the magnitude of the changes are too large to be accounted for by the expected flux uncertainties, which are likely a couple tens of percent, at most.  The fluxes in Table 1 show that the NLR emission increased by a factor of $\sim$2 over $\sim$20 years (1973$-$1989), coincident with the increase in continuum and broad emission-line flux.  It then entered a stable phase for $\sim$1$-$2 decades, serendipitously contemporaneous with the RM campaign monitoring of \citet{Peterson98}.   The relative stability during this period is based on the 2003 SDSS spectrum and 17 spectra taken during JD2448514$-$JD2450122 (1991 Sept.$-$1996 Feb) under photometric conditions.  Emission-line flux measurements made from these 18 spectra demonstrate the rms deviations in the mean [\oiii]\,$\lambda$5007 flux was only 4.3\%.  The flux has since decreased again by a factor of $\sim$2 in the past 10 years.

We see the same general trends in the UV emission lines, with the exception that weak, broad \civ\ and \Lya\ components are still clearly visible in the recent COS spectrum.   Because the BLR is ionization stratified, these emission lines originate nearer the central source than the \Hbeta\ emission that has completely disappeared.  The availability of UV observations is limited by the difficulty of obtaining such data, and as such, the spectra we present are the only available.  It is therefore more difficult to put constraints on (1) the timescales with which the broad UV emission lines have declined, (2) how well they track the decline in optical emission lines, and (3)  whether there truly is an ionization dependence on the remaining presence of broad line emission.  One possible explanation for the remaining broad emission is simply one of equivalent width.  We note that a possible, weak \Halpha\ broad component also remains in our most recent optical spectra.  Nonetheless, we see no evidence of broad \heii\ in the optical ($\lambda$4686) or the UV ($\lambda$1640).  If the ionizing continuum has simply become so weak that the line luminosities have become equivalently weak, then the lower equivalent width lines will `disappear' before the higher equivalent width lines, but a complete BLR (both low and high ionization lines) may still remain intact below our detection limit.

On the other hand, we can put the (near) disappearance of the optical broad lines in context to theoretical models discussed above, if, for instance, the accretion rate of Mrk\,590 has dropped below the limit necessary for the production of a BLR.  We estimate the  bolometric AGN luminosity from the 1450\,\AA\ continuum flux to be $L_{\rm bol}=3.4\times\,10^{42}$\,erg s$^{-1}$, from the 2013 COS spectrum\footnote{This was estimated assuming a bolometric correction at 1450\,\AA\ of 4.20 \citep{Runnoe12, Runnoe12err}}.  This is still orders of magnitude above the advection dominated limit suggested by \citet{Elitzur&Ho09}, which for Mrk\,590 is $L_{\rm bol}\sim\,1.4\times\,10^{40}$\,erg s$^{-1}$, and is also not yet below the limit assumed by \citet{Laor03} of $L_{\rm bol}\sim\,1.4\times\,10^{41}$\,erg s$^{-1}$ for the black hole mass of Mrk\,590.  However, it is somewhat below the limit predicted by \citet{Nicastro00} for the ionizing luminosity expected at the minimum accretion rate necessary to support a disk-wind BLR --- $L_{\rm ion}(\dot{m}_{\rm min}) \sim 2.5\times\,10^{42}$\,erg s$^{-1}$ for the black hole mass of Mrk\,590 --- where, based on our bolometric luminosity and that approximately 24\% of the bolometric luminosity is emitted with wavelengths below 912\AA\footnote{This was estimated using the mean ``Low Luminosity" SED template presented by \citet{Krawczyk13}.}, we find $L_{\rm ion}\sim 8.5\times\,10^{41}$\,erg s$^{-1}$.  Such a result would suggest a physical explanation for the disappearing broad lines if, indeed, the accretion rate has dropped below that necessary for generating a radiation pressure-supported and -driven wind.  However, the reader should keep in mind that these approximations of $L_{\rm bol}$ and $L_{\rm ion}$ have significant uncertainties, so care should be taken in interpreting these results.  Additionally, such interpretations make it difficult to explain the remaining broad components in \civ, \Lya, and \Halpha, if the \Hbeta\ broad line has disappeared as a result of the central engine entering a radiatively inefficient accretion state after this significant decrease in luminosity.  Furthermore, the maximum line width we measure from our data is $\sim$9000\,km s$^{-1}$ for the weak, broad emission component of \civ\ from the 2013 COS spectrum, suggesting that the BLR velocities are nowhere near the $\Delta$v$_{\rm max}\sim$20,000$-$25,000\,km s$^{-1}$ limits discussed by \citet{Nicastro00} and \citet{Laor03}.  Nonetheless, if the luminosity of Mrk\,590 continues its downward trend, it may soon enter these interesting regimes if even the \Halpha\ and UV broad lines continue to disappear.

An additional serendipitous investigation we can perform with the recent UV and optical data (2013 MODS1 and COS spectra) that show only very weak or non-existent broad emission components is to probe NLR line ratios of both the high and low ionization lines.  Many narrow lines are typically `contaminated' by the broad-line emission, such that the similarities and differences between Sy1 and Sy2 line ratios are not well studied, particularly for rest frame UV lines such as \Lya\ and \civ, where not only are observations harder to obtain but absorption is also a problem \citep[see, e.g.,][and references and discussions therein]{Crenshaw&Kraemer05, Stern&Laor13}.  We fit the \Lya\ and \civ\ emission lines in the COS spectrum first with a broad 6th order Gauss-Hermite (GH) polynomial, masking over the narrow line core and any other non-BLR line-specific emission in the profiles (e.g., \Lya\ and \oi\,$\lambda$1301 geocoronal emission, \nv).  We subtracted this component and then fit the residual narrow emission component of each line with a new GH polynomial profile, from which we measure the presumed NLR contribution to the flux.  The broad and total profile models are shown in red in Figure \ref{fig_uvspec} for each of these lines.  We find the following observed line strengths, relative to narrow \Hbeta, for these and other common narrow emission lines in the optical/UV:  \Lya\ 40.26; \civ\ 13.50; \heii\,$\lambda$1640 0.23; [\oiii]\,$\lambda$5007 12.52; \Halpha\ 2.75.  Rest-frame UV narrow-line components were detected and ratios reported for one other Type 1 AGN, 3C\,390.3 \citep{Ferland79}, but given the lower resolution and S/N of that data and the fact that narrow-line fluxes in this object may be affected by jet effects \citep{Zheng95, Zheng96}, a comparison with the present results is of questionable merit.  On the other hand, our measured values are very similar to the relative intensities given by \citet{Ferland&Osterbrock86} for Sy2 spectra.  This could indicate that the NLR line ratios of Sy1s and Sy2s are intrinsically similar.  However, it could instead be possible that Sy1 and Sy2 ratios are intrinsically different, but by virtue of the transition Mrk\,590 is experiencing, we have simply recovered the ratios expected for the state that Mrk\,590 appears to be subsuming, i.e., a Sy2.

A final note of interest is simply that we detect a significant narrow emission component of \civ.  There has been extensive discussion in the literature about whether a ``true'' narrow, or more accurately, an NLR component of \civ\ exists (i.e., that emitted co-spatially with forbidden lines from a low-density, very spatially extended gas distribution). This is of particular interest for estimating black hole masses based on \civ, as this affects the necessity (or not) for subtraction of such a component from the observed \civ\ profile \citep[e.g., see discussions by][and references therein]{Baskin&Laor05, Vestergaard06, Sulentic07, Shen12, Denney12}.  The COS spectrum in Figure 2 clearly shows such a component in Mrk\,590, and the line strengths given in Table 1 show it to be as strong as [\oiii]\,$\lambda$5007.  The FWHM of this and other narrow emission line components from the optical/UV spectra, also given in Table 1, demonstrate that while this \civ\ component is ``narrow'', neither the \civ\ nor \Lya\ integrated narrow emission-line flux are being emitted co-spatially with the integrated emission from the other narrow forbidden or Balmer lines.  Instead, taking the velocity widths as virial velocities suggests the narrow \civ-emitting gas resides roughly a factor of 3 closer to the nucleus than that emitting [\oiii]\,$\lambda$5007.  This supports the argument that such emission is coming from an ``inner'' narrow line region (see e.g., \citealt{Nelson00}, \citealt{Kraemer00}, and discussion by \citealt{Denney12}) and should therefore be removed prior to BLR line width measurements.  However, defining and separating such a contribution to a \civ\ profile such as that visible in the Mrk\,590 IUE spectra, for example, has historically been the difficulty.  Therefore, reliably subtracting it in the presence of a strong broad emission component without an unblended template for the velocity width, which does not exist, remains problematic.  The closest proxy would be using the approach of Crenshaw et al.\ \citep[e.g.,][]{Crenshaw&Kraemer07}, who have previously used the narrow component of \heii\,$\lambda$1640 as a template for a narrow \civ\ component.  Our narrow line-width measurements here suggest this approach is better than using a forbidden narrow emission line, but we still find the FWHM of \heii\ to be narrower than that of \civ.  Furthermore, Crenshaw et al.\ have typically used this method with very high quality (typically low redshift, space-based) data; however, the intensity of narrow \heii\ is much weaker than that of \civ\ (by a factor of $\sim$150), so application of this method to high-redshift AGN spectra, where assessing the uncontaminated BLR velocity from \civ\ is most desirable for black hole mass estimates, would be, unfortunately, tenuous at best. 

\section{Discussion}
\label{S_discuss}

The seeming ability of objects like Mrk\,590 to transition between Type 1 and Type 2 without an otherwise obvious flaring event like what \citet{Shappee13} observed in NGC\,2617 raises the question of what assigning a type is really saying about an AGN.  It is possible that such classifications should be taken as more of a statement of current `state' rather than a true, fixed `type', as Mrk\, 590 further indicates that type can change.  This distinction may be of particular importance since it is becoming apparent that whether a particular AGN appears at any given time to be Type 1 or Type 2 (or somewhere in between) may be saying something more significant about the AGN physics and the immediate environment of the black hole at that particular time than strictly about the viewing angle.  For Mrk\,590, the decrease in the BLR, the UV, and the X-ray emission all seem to point to a decrease in total luminosity, thus accretion rate, and not simply a change in obscuration.  Furthermore, the dynamical timescale of the \Hbeta\ emitting region is roughly 8 yrs in the rest frame of Mrk\,590, based on the black hole mass and \Hbeta\ FWHM given by \citet{Peterson04}.  An optically thick medium capable of occulting the BLR could only reside outside the dust sublimation radius and would therefore have a much longer dynamical timescale.  Yet, the broad \Hbeta\ emission transformed from strong to non-existent in $\lesssim$10 years (1996--2006).  Furthermore, we see a dramatic change in the strength of not only the BLR emission but also the continuum and the NLR emission.  If obscuration were responsible for this, the obscuring medium would need to cover both our line of sight to the continuum source, in which case we may expect to see evidence of an obscurer in our X-ray continuum model, which we do not, as well as that between the continuum and the NLR, which is highly improbable. Therefore, though it cannot be definitively ruled out, line of sight obscuration is an unlikely explanation for the observations we present here.   In general, the acquisition of sufficient data to determine the bolometric luminosity and spectral energy distribution of an AGN before and after a change in type should be able to distinguish between these two physical origins for the change. Changes in accretion rate should affect the flux at all wavelengths.  However, flux changes due to obscuration will be wavelength dependent, where the decrease in UV/optical emission will be countered by an increase in the mid-to-far IR flux, since the obscuring medium absorbs the higher energy photons but then reradiates them as thermal emission at these long wavelengths.

A more likely explanation for the suite of observations presented here is that an accretion event occurred 40$+$ years ago.  This event was capable of stimulating the production of ionizing photons that have excited the BLR emission over the subsequent 40 years.  However, the energy produced in such an event has now been depleted.  The symmetry of the narrow emission lines and the time scales over which we see significant changes in both the broad and narrow emission lines are too short for it to be plausible that this accretion event actually created these regions, i.e., to initially form and light up the AGN from a quiescent state.  Instead, it is likely that at least the bulk of the nuclear gas was already present, just not emitting significant broad-line emission.  This suggests that the appearance of a BLR may be episodic over the accretion history of supermassive black holes, depending on the reservoir of material available in the nucleus for accretion.  It further demonstrates the importance of repeat observations of larger samples of Seyfert galaxies over longer baselines as a means to better quantify how often this behavior may occur.  Transition objects like Mrk\,590 may also be a link to other `abnormal' AGN found from large survey samples.  For example, \citet{Roig14} found unusual broad-line objects among the BOSS sample of luminous galaxies that show a broad \mgii\ emission line component but little to no broad Balmer emission.  An interesting possibility is that these objects are in a similar transition state as Mrk\,590.  Unfortunately, a recent spectrum of the \mgii\ region in Mrk\,590 has not yet been obtained, so more data are needed to connect the likely time variable nature of the broad Balmer lines in Mrk\,590 with this larger class of objects.  We have been awarded time to obtain a spectrum of the \mgii\ region in Mrk\,590 with \HST\ in Cycle 21 for this purpose and will therefore be able to address this in future work.

Under the assumption that the observed behavior of Mrk\,590 {\it was} the consequence of a past accretion event, we can infer additional physical properties of this system.  The BLR would have known and responded to a new accretion event almost instantaneously, given the \Hbeta\ radius of $\sim$10$-$20 light days \citep{Peterson04} and the short recombination time in this region.  However, this is not the case for gas at the distance and densities of the NLR.  The behavior and subsequent properties of the \ob\ lines, as well as the heterogeneous data in our sample, allow us to put various independent limits on the NLR radius. One of the weakest, but nonetheless most reliable constraints on the NLR size comes from the fact that we measure nearly identical [\oiii] fluxes from the 2003 SDSS spectrum as we do from the large aperture RM campaign spectra taken in the 1990's.  This indicates that most of the [\oiii] emission must arise from the spatially unresolved region with a 1.7\,kpc diameter --- the physical size subtended by the SDSS aperture at the distance of Mrk\,590.   Next, insofar as we can rely on the observed increase of the [\oiii]\,$\lambda$5007 flux by 40\% between 1973 and 1983, light travel time arguments suggest that the NLR emission must be $\lesssim3$\,pc from the central source.  In addition, the fluxes measured from the 2003 SDSS spectrum show that the continuum and BLR emission had significantly declined already, yet the [\oiii] flux has not yet changed.  However, the [OIII] flux {\it had} diminished by the time the 2006 MDM spectrum was obtained (again assuming at least some amount of trust in this flux measurement).  Thus, even if this decline began right after the final RM campaign [OIII] flux calibration spectrum was obtained, we deduce a light travel timescale of $\lesssim$10 years, consistent with the distance inferred from the earlier increase in flux.   

Next, we briefly consider the velocities of the [\oiii]-emitting gas, though the evidence for such a discussion is less robust than that derived from the fluxes.  This is because large ($\sim$100 \kms) systematic errors in the velocity widths derived from individual spectra taken through large spectroscopic apertures are possible.  This is on account of the difficulty involved in accurately measuring and correcting for the spectral resolution given that the PSF does not fill the slit. \citet{Vrtilek&Carleton85} measure the [\oiii]\,$\lambda$5007 velocity width to be 397$\pm$22\,\kms\ from a high (23\,\kms) resolution spectrum taken sometime between 1980 June and 1981 July.  This is consistent with our resolution corrected velocities, for the most part.  If we at least assume that these velocities approximately represent the virial velocity at the radius of the NLR, this also indicates consistent distances with other limits we considered above, with velocities $\sim$250$-$550\,km s$^{-1}$ corresponding to distances of $\sim$0.7$-$3.3\,pc.  While we regrettably cannot trust the reliability of the 1973 line width measurement, it is notably larger than the other measurements.  This could lead to the interesting, though highly speculative, consideration that with this early observation, we are actually observing the NLR `lighting up' from the inside out as the AGN continuum luminosity begins to increase significantly.  Assuming a 10$^4$ K gas, the recombination time for gas near the critical density of \ob\ will be on the order of days; however, as the density drops with increasing radius to the expected average density of the NLR of $\sim$2000 cm$^{-3}$ \citep{Koski78}, the recombination time approaches a decade.  Such effects could lead to a scenario in which we see narrow-line emission only from the higher-velocity, more dense gas in the innermost regions of the NLR as the AGN continuum first brightens, and only after years will the velocity widths of the integrated [\oiii]-emitting gas decrease as the spatial extent of the emission increases.  Certainly, the range of velocities measured from the narrow component of the various emission lines we consider, which are significantly larger than the line width uncertainties, indicate that ``narrow'' emission line gas is emitted from an extended spatial region with a density gradient.  The innermost emission from the high ionization species, such as \civ, \Lya, and \heii\ is coming from well within sub-pc scales.
 
Finally, to complete the picture of events, our observations indicate that as the AGN continuum luminosity increased (decreased), the BLR emission strengthened (diminished) faster than the NLR emission.  This behavior suggests the spatial propagation of information from the central source outward because of density effects and light travel time.  This again implies that an ``event" of some sort may have triggered the creation of enough ionizing photons to excite the observed changes in BLR emission and subsequent strengthening of the NLR emission (as discussed above).  There is nothing in the observations that can yet point to ``what" may have been accreted to cause the observed sequence of events.  However, we can {\it roughly} estimate the total mass needed to power the bolometric luminosity integrated over the span of our observations with a simple, back of the envelope calculation.  We take the 5100\,\AA\ fluxes from all optical spectra shown in Figure~\ref{fig_allspec} (values in Table 1) and convert these to bolometric luminosities assuming a bolometric correction of 8.10 \citep{Runnoe12err}.  We then estimate the integrated luminosity both under the rough continuum emission light curve shown in Figure \ref{fig_lightcurves} and by extending the observed rise and decline, modeled as Gaussian wings, to both the past and future to conservatively cover a temporal range from 1950 to 2020.  Finally, assuming that the mass-to-energy conversion efficiency is 7\%, we find that the observed continuum output during this event can be accounted for with only $\sim$1$-$2\,$M_{\odot}$ of total mass being accreted over these 70 years.  
 
\section{Summary}
\label{S_summary}

We have presented optical, UV, and X-ray observations of the `Classical' Seyfert 1 Mrk\,590 that span the past 40+ years.  This interesting object brightened by a factor of 10's between the 1970's and 1990's and then faded by a factor of a 100 or more at all continuum wavelengths between the mid-1990's and present day.  Notably, there is no evidence in the current data set that this recent, significant decline in flux is due to obscuration; in particular, the most recent X-ray observations are consistent with zero intrinsic absorption.  There were similarly dramatic changes in the emission-line fluxes.  The most striking change is the complete disappearance of the broad component of the \Hbeta\ emission line, which had previously been strong (equivalent widths $\sim$20$-$60 \AA) and the focus of a successful reverberation mapping campaign that resulted in a secure estimate of the supermassive black hole contained within the nucleus of Mrk\,590 of $\sim5\times10^7 M_{\odot}$.  As a result of these significant changes, the optical spectrum of Mrk\,590 currently looks more like that of a Sy2 AGN, with predominantly only narrow emission lines and a strong host galaxy stellar continuum.

The changes in emission line properties over this time period allowed us to determine that Mrk\,590, at least in its current state, has NLR emission line ratios that are similar to those measured in Sy2 spectra.  We were also able to put limits on the radius of the [\oiii]-emitting region of the NLR to be $\sim$0.7$-$3.3\,pc based on the observed changes in the integrated line flux and the velocity width of [\oiii]\,$\lambda$5007.  These results are consistent with the NLR size determined by \citet{Peterson13} for NGC\,5548, another nearby Sy1 galaxy containing a BH of similar mass to Mrk\,590.

The implications arising from this long time series of Mrk\, 590 are that (1) Mrk\,590 is a direct challenge to the historical paradigm that AGN type is exclusively a geometrical effect, and (2) there may not be a strict, one-way evolution from Type 1 to Type 1.5$-$1.9 to Type 2 as recently suggested by \citet{Elitzur14}.  Instead, for at least some objects, the presence of BLR emission may coincide only with episodic accretion events throughout a single active phase of an AGN.  If true, such behavior may be more prominent in Seyfert galaxies, where accretion has been theorized to be a consequence of secular processes and therefore likely more episodic than quasar activity, which may be triggered more predominantly by major mergers \citep[see, e.g.,][]{Sanders88, Treister12}.  The final possibility, of course, is that we are witnessing the final stages in the life of this AGN, and Mrk\,590 is completely turning off.  This would be an incredible find, but is likely the most improbable explanation, given the low duty cycles for low redshift AGN \citep[see, e.g.,][]{Schulze&Wisotzki10, Shankar13}. We plan to continue monitoring Mrk\,590 to look for further changes in its current state, but only time and more observations will tell whether the BLR returns.

\acknowledgements Support for program number GO-13185 was provided by NASA through a grant from the Space Telescope Science Institute, which is operated by the Association of Universities for Research in Astronomy, Inc., under NASA contract NAS5-26555. KDD is supported by an NSF AAPF fellowship awarded under NSF grant AST-1302093. BMP and GDR are grateful for NSF support through grant AST-1008882 to The Ohio State University.  MCB acknowledges the NSF for support through the CAREER Grant AST-1253702 to Georgia State University.  This paper uses data taken with the MODS spectrographs built with funding from NSF grant AST-9987045 and the NSF Telescope System Instrumentation Program (TSIP), with additional funds from the Ohio Board of Regents and the Ohio State University Office of Research.  This work was based in part on observations made with the Large Binocular Telescope.  The LBT is an international collaboration among institutions in the United States, Italy and Germany. The LBT Corporation partners are: the University of Arizona on behalf of the Arizona university system; the Istituto Nazionale di Astrofisica, Italy; the LBT Beteiligungsgesellschaft, Germany, representing the Max Planck Society, the Astrophysical Institute Potsdam, and Heidelberg University; the Ohio State University; and the Research Corporation, on behalf of the University of Notre Dame, the University of Minnesota, and the University of Virginia.


\begin{deluxetable*}{lcrcccc}
\tablecolumns{7}
\tablewidth{0pt}
\tablecaption{Mrk\, 590 Multi-wavelength Spectral Properties}
\tabletypesize{\scriptsize}
\tablehead{
\colhead{Observation} &  \colhead{Year of} & \colhead{Continuum} & \colhead{Continuum} & \colhead{Narrow Line} & \colhead{Narrow Line} & \colhead{Narrow Line}\\
\colhead{Identifier}&\colhead{Observation} & \colhead{Region\tablenotemark{a}} & \colhead{Flux\tablenotemark{b}}&\colhead{Identifier} &\colhead{Flux\tablenotemark{b}} &\colhead{FWHM\tablenotemark{b}}\\
\colhead{(1)} &
\colhead{(2)} &
\colhead{(3)} &
\colhead{(4)} &
\colhead{(5)} &
\colhead{(6)} &
\colhead{(7)}
}


\startdata
{\bf Optical Spectra}  &              &			&	      &              		       &                  &                \\
Lick IDS			& 1973	&5100\,\AA\ 	& 3.4 \phantom \phantom \phantom    & [\oiii]\,$\lambda$5007 & 0.52 & 584 km s$^{-1}$ \\
Perkins OSU IDS	& 1983	&5100\,\AA\ 	& 11.   \phantom  & [\oiii]\,$\lambda$5007 & 0.70 & 390 km s$^{-1}$ \\
RM campaign		& 1989	&5100\,\AA\ 	& 55.     & [\oiii]\,$\lambda$5007 & 1.1 & 246 km s$^{-1}$ \\
RM campaign		& 1993	&5100\,\AA\ 	& 26.     & [\oiii]\,$\lambda$5007 & 1.1 & 327 km s$^{-1}$ \\
RM campaign		& 1996	&5100\,\AA\ 	& 42.     & [\oiii]\,$\lambda$5007 & 1.1 & 227 km s$^{-1}$ \\
SDSS 			& 2003	&5100\,\AA\ 	& 1.3     & [\oiii]\,$\lambda$5007 & 1.1 & 402 km s$^{-1}$ \\
MDM			& 2006	&5100\,\AA\ 	& 0.28  \phantom & [\oiii]\,$\lambda$5007 & 0.87 & 479 km s$^{-1}$ \\
LBT MODS1		& 2013	&5100\,\AA\ 	& $<$0.014 & [\oiii]\,$\lambda$5007 & 0.42 & 448 km s$^{-1}$ \\
LBT MODS1		& 2013	&5100\,\AA\ 	& $<$0.014 & \Hbeta\ (narrow)	& 0.033 & 425 km s$^{-1}$ \\
LBT MODS1		& 2013	&5100\,\AA\ 	& $<$0.014 & \Halpha\ (narrow)	& 0.120 & 457 km s$^{-1}$ \\
KOSMOS			& 2013	&5100\,\AA\ 	& $<$0.10   & [\oiii]\,$\lambda$5007 & 0.47 & 399 km s$^{-1}$ \\
MDM			& 2014	&5100\,\AA\ 	& $<$0.11   & [\oiii]\,$\lambda$5007 & 0.56 & 338 km s$^{-1}$ \\
\hline
{\bf UV Spectra}        &              &			&	           &              			&               &                           \\
IUE				& 1982	&1450\,\AA\	& 88.5   &  \nodata			& \nodata  & \nodata \\
IUE				& 1991	&1450\,\AA\	& 388.   &  \nodata			& \nodata	& \nodata \\
COS				& 2013	&1450\,\AA\	& 3.7	     & \CIV\	 (narrow)	        & 0.45	& 939 km s$^{-1}$ \\
COS				& 2013	&1450\,\AA\	& 3.7	     & \Lya\	 (narrow)		& 1.3		& 845 km s$^{-1}$ \\
COS				& 2013	&1450\,\AA\	& 3.7	     & \heii\,$\lambda$1640 (narrow) & 0.0075 & 612 km s$^{-1}$ \\
\hline
{\bf X-ray Spectra\tablenotemark{c}}     &             &     			&           &              			&               &                           \\
EXOSAT			& 1984  	&2$-$10\,keV	& 27.	$\pm$2.7     & \nodata			& \nodata	 & \nodata	\\
RASS			& 1991	&0.1$-$2.4\,keV & 46.3$\pm$28.7   & \nodata			& \nodata 	 & \nodata \\
XMM				& 2004	&0.2$-$2\,keV & 3.31$\pm$0.11 & \nodata			& \nodata   & \nodata \\
XMM				& 2004	&2$-$10\,keV  	& 6.95$\pm$0.11 & \nodata			& \nodata	  & \nodata \\
Chandra			& 2004	&$0.5-10$\,keV	& 11.9$\pm$0.3 & \nodata			& \nodata	   & \nodata \\
Suzaku			& 2011	&2$-$10\,keV & 6.8 & \nodata				& \nodata	 & \nodata \\
Chandra			& 2013	&$0.5-10$keV	& 1.3$\pm$0.1 & \nodata				& \nodata	 & \nodata 
\enddata

\tablenotetext{a}{UV and optical continuum regions refer to rest frame wavelengths, but the energy range of the X-ray observations is that observed.}  
\tablenotetext{b}{UV and optical continuum fluxes are in units of 10$^{-16}$ erg s$^{-1}$ cm$^{-2}$ \AA$^{-1}$, X-ray fluxes are in units of 10$^{-12}$ erg s$^{-1}$ cm$^{-2}$, and emission line fluxes are in units of 10$^{-13}$ erg s$^{-1}$ cm$^{-2}$.   We have not included uncertainties in the optical/UV flux or line width measurements because the data from which these quantities are measured are typically very high S/N, so that uncertainties in these quantities will be dominated by unmeasurable systematics present in this extremely heterogeneous data set, rather than by observational or statistical sources.}
\tablenotetext{c}{References for past X-ray data include EXOSAT: \citet{Turner89}; RASS: \citet{Voges99, Mahony10}; XMM: \citet{Longinotti07, Bianchi09};  Chandra 2004: \citet{Longinotti07}; Suzaku: \citet{Rivers12}.} 

\label{Tab_fluxes}
\end{deluxetable*}


\end{document}